\documentclass{PoS}

\title{The QCD EoS from simulations on BlueGene L Supercomputers at LLNL and NYBlue}

\ShortTitle{EoS determination}

\author{
  \speaker{Rajan Gupta} 
  (HotQCD Collaboration\thanks{
  Alexei Bazavov,
  Tanmoy Bhattacharya,
  Michael Cheng,
  Norman Christ,
  Carleton DeTar,
  Steven Gottlieb,
  Rajan Gupta,
  Urs Heller,
  Kay Huebner,
  Chulwoo Jung,
  Frithjof Karsch,
  Edwin Laermann,
  Ludmila Levkova,
  Thomas Luu,
  Robert Mawhinney,
  Peter Petreczky,
  Dwight Renfrew,
  Christian Schmidt,
  Ron Soltz,
  Wolfgang Soeldner,
  Robert Sugar,
  Doug Toussaint,
  and
  Pavlos Vranas
  }
  ) \\
  Theoretical Division, Los Alamos National Laboratory, Los Alamos, NM 87545, USA\\
  E-mail: \email{rajan@lanl.gov}
}

\newcommand{\pbp}{\langle \bar \psi \psi \rangle}

\abstract{We present results for the QCD Equation of State (EoS)
  obtained using simulations of lattice QCD at zero chemical
  potential.  Our high statistics results compare improved asqtad and
  p4fat3 staggered quarks on lattices with a temporal extent $N_\tau =
  6$ and $8$ and light quark masses approximately one fifth and one
  tenth the strange quark mass.  We find that the two actions give
  consistent results and estimate that the trace anomaly $(\varepsilon
  - 3p)/T^4$ obtained on $N_\tau = 8$ lattices represents the
  continuum value to better than $20\%$ uncertainty over the
  temperature range $140-700$ MeV.  The precision in the 
  estimates of energy density and pressure are better, therefore, we 
  conclude that lattice estimates of the energy density and pressure
  should be used in the phenomenological analysis of RHIC and
  LHC data. We also find a consistent picture of the
  crossover temperature from all observables studied, with the best
  estimated range to be $185-195$ MeV.  These calculations are being
  carried out on the IBM BlueGene/L supercomputer at Lawrence
  Livermore National Laboratory and at the New York Center for
  Computational Science (NYBlue).}

\FullConference{The XXVI International Symposium on Lattice Field Theory\\
   July 14-19 2008\\
    Williamsburg, Virginia, USA}

\begin{document}

\section{Introduction}

Experiments at the Brookhaven Relativistic Heavy Ion Collider (RHIC),
and at the LHC, will study the formation and evolution of the quark
gluon plasma.  The phenomenological analysis of the data has relied on
a hydrodynamic description of the medium, and the results suggest a
strongly interacting plasma with very small viscosity. A crucial
input (and eventually the goal) of this phenomenological analysis
is the characterization of the equation of state (EoS) over the temperature
range $140-700$ MeV relevant to experiments at both RHIC and
LHC. Theoretically, simulations of Lattice QCD provide a first
principles analysis of QCD in the vicinity of thermal equilibrium and
zero baryon and strangeness chemical potential (For recent reviews
see~\cite{Detar:2008,Karsch:2007}). The results of our detailed
analysis of the equation of state as a function of temperature will,
therefore, provide crucial guidance in the phenomenological
interpretation of experimental measurements.

A related quantity of high interest is the transition temperature from
hadrons to a quark-gluon plasma (QGP). The minimum energy density required
to produce a quark-gluon plasma grows as the fourth power of the
temperature.  Thus, a 10\% error in the threshold temperature
corresponds to a 45\% error in the threshold energy density.  Previous
calculations with staggered fermions have found the transition for
$2+1$ flavors to be a rapid crossover. In light of this lack of a true
phase transition (see, for example, \cite{Aoki:2006br}), we will
discuss what quantities to focus on for estimating the transition
temperature needed in the phenomenological analysis of heavy ion
collisions.

This talk updates the status of HotQCD results for the EoS and the
transition temperature obtained from large scale simulations on the
BlueGene L at LLNL and NYCCS
(NYBlue)~\cite{DetarGupta:2007,Detar:2008,Karsch:2007}. These
calculations have been carried out with two sets of ${\cal O}(a^2)$
improved actions, asqtad and p4fat3 staggered fermions, and for each
of these actions we have simulated on lattices with extent $N_\tau =
6$ and $8$ in the Euclidian time direction and a spatial extent of
$32^3$. The strange quark mass is fixed to roughly the physical
strange quark value, and $m_l/m_s = 0.1$ and $0.2$ corresponding to
Goldstone $M_\pi \approx 215$ and $304$ MeV. For preliminary results
at $m_\ell = 0.05 m_s$ see talk by Soeldner~\cite{Soeldner:2008}. The
associated $T=0$ calculations needed to subtract ultraviolet
divergences in the EoS and determine the lattice spacing were done on
$32^4$ or larger lattices. We discuss uncertainties associated with
the continuum and chiral extrapolations based on these combined data
sets.

\section{Parameter Sets Used in the Simulations}

The run parameters for simulations with the p4fat3 action are given in
Table~\ref{tab:p4param}, and those for the asqtad action in
Table~\ref{tab:asqtadparam}. The lattice scale is set using $r_0$ (or
equivalently $r_1$ as they give consistent estimates). The simulations
are being carried out along lines of constant physics by adjusting the
bare strange quark mass to produce an approximately constant physical
value of $M_{\bar ss} = 686$ MeV corresponding to $M_{\bar ss}r_0 =
1.58$ and $M_\pi r_0 \approx 0.52$ for $M_\ell=0.1m_s$ along the
trajectory. Having fixed the strange quark mass we are simulating
three light quark mass values held fixed at $m_\ell/m_s = 0.2$, $0.1$
and $0.05$. To allow comparison with previous studies the asqtad trajectory 
in HotQCD simulations is set 20\% higher than the physical strange quark mass.

\begin{table}[ht]
  \begin{center}
\begin{tabular}{llllll}
$N_\tau$  & $N_s$ & $m_\ell/m_s$ &  Total \#  of   & Independent    & Total \# trajectories  \\
          &       &              &  $\beta$ values  &    Streams     & per $\beta$ (0.5 length) \\
\hline\hline
$6$       &  $16$      &  $0.2$    &  $9$      &  RBC-Bielefeld       &  10-60K  \\
\hline
$6$       &  $16$      &  $0.1$    &  $8$      &  RBC-Bielefeld       &  25-60K  \\
          &  $24$      &           &           &  Collaboration~\cite{Cheng:2006}       &    5-8K    \\
 \hline
$8$       &  $32$      &  $0.2$    &  $7$      &  $2$       &  14-15K  \\
\hline
$8$       &  $32$      &  $0.1$    &  $30$     &  $1-3$     &  8-37K  \\
\hline
$32$     &  $32$       &  $0.1$    &  $21$     &  $1$       &  2-6K  \\
\end{tabular}
  \end{center}
\caption{Simulation parameters for the p4fat3 action. For thermalization
  800 trajectories are discarded. We also indicate the number of independent 
  streams used to accumulate statistics. }
   \label{tab:p4param}
\end{table}
\begin{table}[ht]
  \begin{center}
\begin{tabular}{llllll}
$N_\tau$  & $N_s$ & $m_\ell/m_s$ &  Total \# of     & Independent    & Total \# trajectories  \\
          &       &              &  $\beta$ values  &    Streams     &    per $\beta$ (unit length) \\
\hline\hline
$6$       &  $32$   &  $0.2$    &  $7$     &  $1$        &  18-20K  \\
\hline
$6$       &  $32$   &  $0.1$    &  $7$     &  $1$        &  18-19K  \\
\hline
$8$       &  $32$   &  $0.2$    &  $8$     &  $1$        &  13K  \\
\hline
$8$       &  $32$   &  $0.1$    &  $23$    &  $1$        &  15-16K  \\
\hline
$8$       &  $64$   &  $0.1$    &  $1$     &  $1$        &  3.7K  \\
\hline
$32$      &  $32$   &  $0.1$    &  $18$    &  $1$        &  5-6K  \\
\end{tabular}
  \end{center}
\caption{Simulation parameters for the asqtad action. 
1000-1200 time units are discarded for thermalization.}
  \label{tab:asqtadparam}
\end{table}

\section{Strategy for Precision Calculations}

Our goal is to map out the EoS over the temperature range 140-700 MeV
to within $5\%$ uncertainty. In addition to performing high statistics
simulations (see Tables~\ref{tab:p4param} and~\ref{tab:asqtadparam})
we have adopted the following strategy to understand and control
systematic errors.

\begin{enumerate}
  \item Discretization Errors: We are simulating two improved staggered 
    actions -- asqtad and p4fat3 -- that have different $O(a^2)$ errors. 
    This provides a check but does not address the rooting issue for staggered fermions. 
    \label{item:actions}
  \item	Continuum Limit: Our goal is to perform a continuum extrapolation along lines of 
    constant physics using lattices with $N_\tau = 6$, $8$ and $12$. 
    \label{item:continuum}
  \item Extrapolation to physical $u,d$ quark masses and the chiral
    limit will be done using $m_\ell/m_s$ = $0.2$, $0.1$ and
    $0.05$. Staggered taste violations in finite $T$ calculations 
    require resolving what effective $M_\pi$ should correspond to these 
    quark masses and thus the physical value of $m_{u,d}$~\cite{Detar:2008}.
  \item Crossover Temperature: Recognizing the absence of a phase transition at 
    physical values of the quark masses, we are simulating at 
    $2-5$ MeV interval over the entire crossover region to provide a precise 
    quantitative picture of the transition in energy density, pressure, etc.
  \item $T=0$ simulations needed for performing subtractions of
    lattice artifacts in the determination of the EoS: We are
    simulating almost as many $\beta$ values ($\approx 20$ for each
    action at $N_\tau=8$) as used in finite temperature
    runs. Estimates of $a$ includes the much larger asqtad
    zero-temperature program.
    \label{item:zerotemp}
\end{enumerate}

\section{Equation of State}

\begin{figure}[ht]
  \begin{tabular}{cc}
    \includegraphics[width=.5\textwidth]{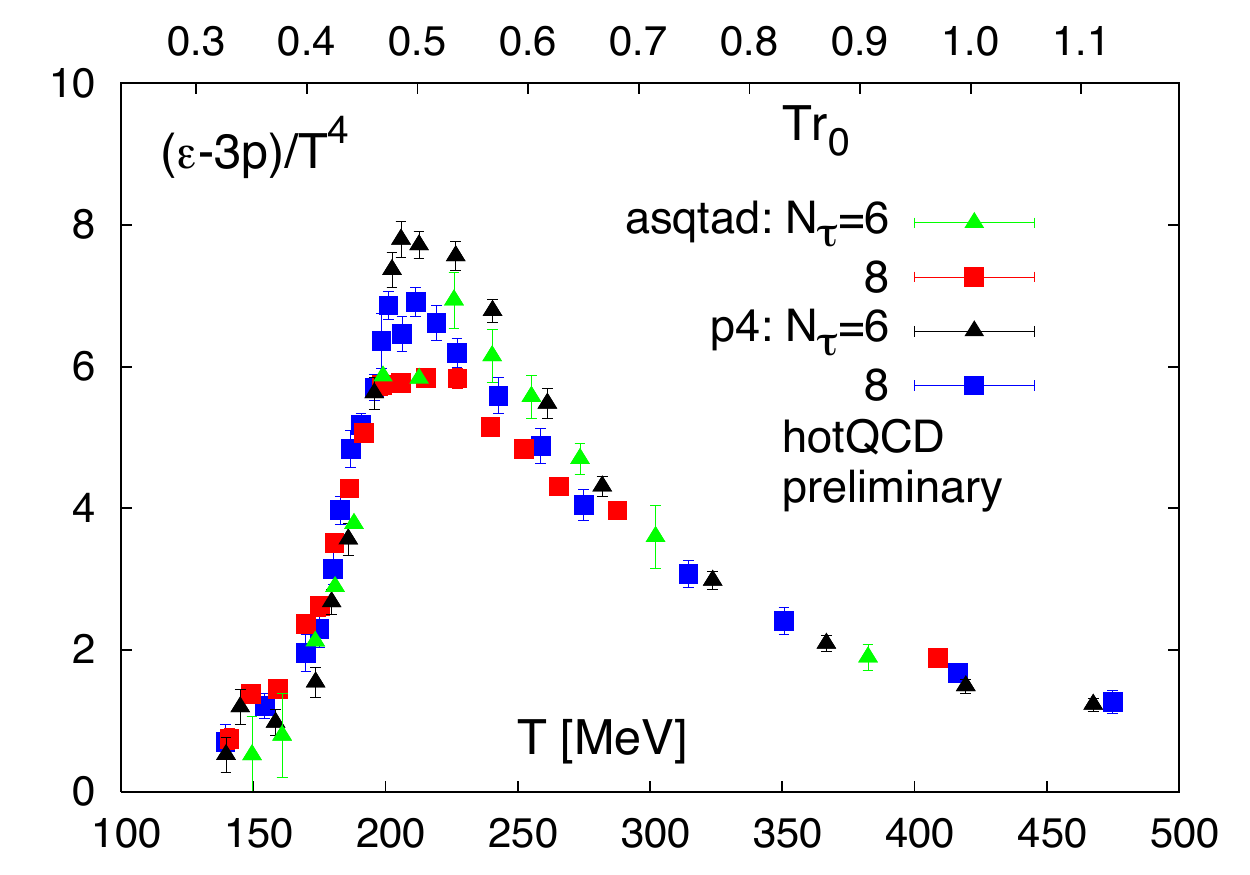}
    &
    \includegraphics[width=.5\textwidth]{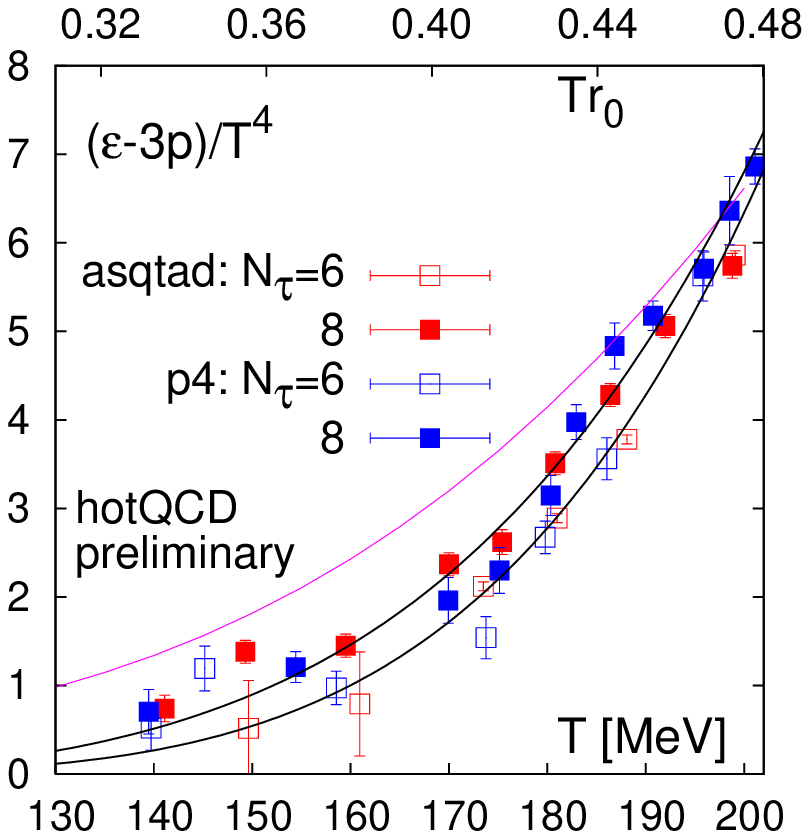}
  \end{tabular}
  \caption{Preliminary results for the trace anomaly for the p4fat3 and asqtad
   actions as a function of temperature in MeV and in units of $r_0$.
   The figure on the right magnifies the range $T=140-200$ MeV and includes
   fits to the p4fat3 points. 
   The purple curve above the data is from the hadron resonance gas model.  
  \label{fig:EoS}
  }
\end{figure}

\begin{figure}[ht]
  \begin{tabular}{cc}
    \includegraphics[width=.5\textwidth]{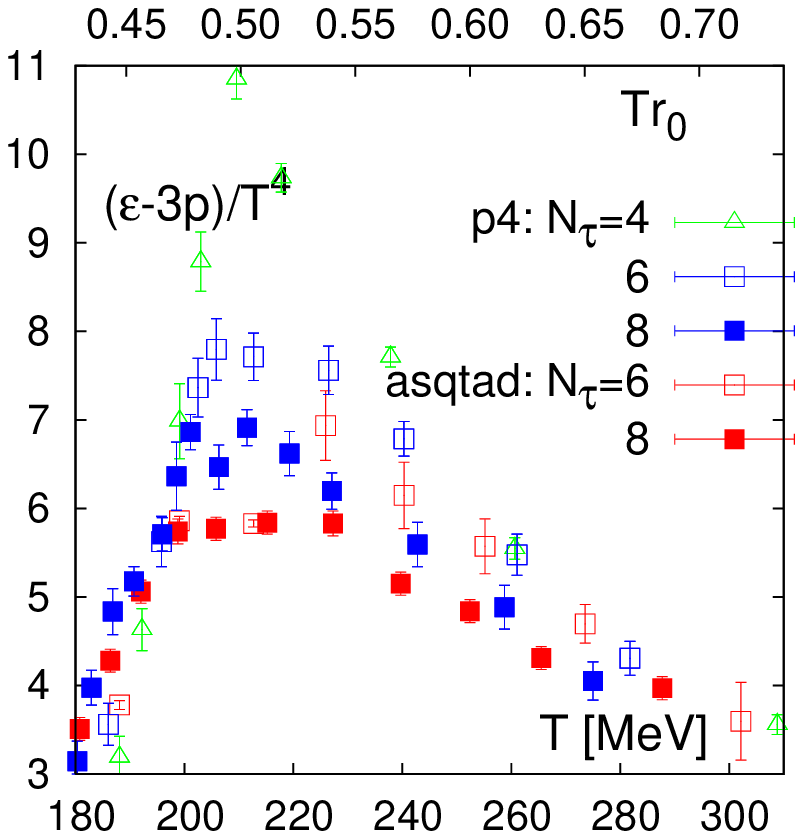}
    &
    \includegraphics[width=.5\textwidth]{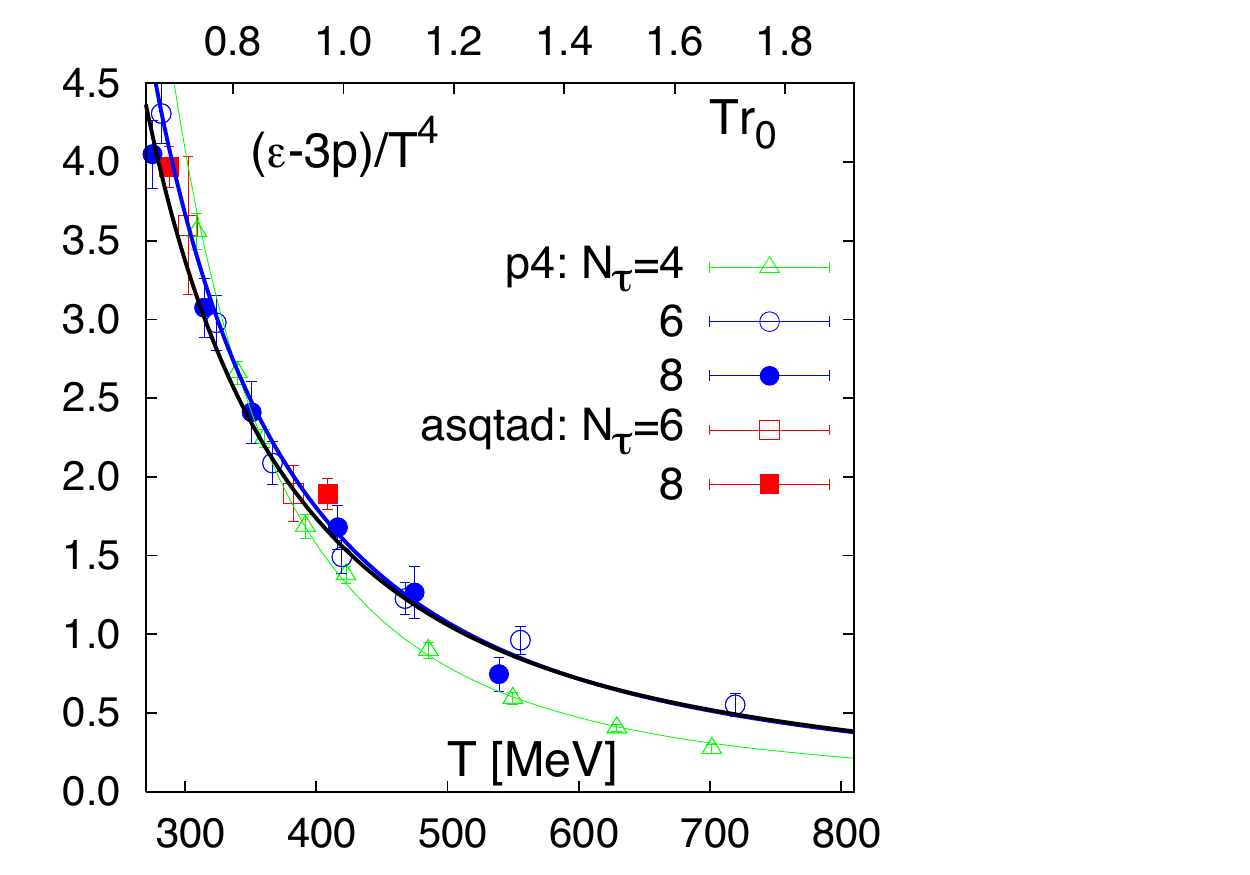}
  \end{tabular}
  \caption{Details of EoS data for the p4fat3 and asqtad
   actions. The left figure shows the range $T=180-300$ MeV
   and the right figure shows the range $T=300-700$ MeV and fits to the p4fat3 data.
  \label{fig:EoSdetail}
  }
\end{figure}

Our preliminary results for the trace anomaly ($(\varepsilon - 3p)/T^4$) for both
actions and for $N_\tau=6$ and $8$ lattices are shown in
Fig.~\ref{fig:EoS}. Over the full range $T=140-500$ MeV, we find the
same pattern for both actions on going from $N_\tau=6$ to $8$, $i.e.$
a decrease in peak height and a slight shift of points to lower $T$,
with the largest change in the range $T=190-300$ MeV.  A more detailed
picture of the data over $(140 \leq T \leq 200)$ MeV is also presented
in Fig.~\ref{fig:EoS} (right figure), and in Fig.~\ref{fig:EoSdetail} for $(180 \leq T
\leq 300)$ MeV (left figure), and $(300 \leq T \leq 700)$ MeV (right figure).


Data in the range $T=140-200$ MeV lie below the hadron resonance gas
(HRG) model.  There is an upward shift (larger value of $(\varepsilon - 3p)$)
on going from $N_\tau=6$ to $8$, which for some of the points is comparable
to the difference between the $N_\tau=8$ data and the HRG estimate.

The discretization effects are most pronounced in the range
$T=180-300$ MeV. We find up to $20\%$ decrease in the peak on going
from $N_\tau=6$ to $8$ for both actions and the asqtad data lie below
p4fat3 values by up to $15\%$. We find that the position of the peak
remains above $200$ MeV, as already observed in the 
$N_\tau=4$ data with p4fat3 action~\cite{Karsch:2007,DetarGupta:2007}. 

At higher temperatures ($T > 300$ MeV) the $N_\tau =6$ and $8$ data
are consistent. There, however, are two unresolved issues. The first
is possible finite volume effects at $T>400$ MeV. The one $N_\tau=8$
asqtad point at $T=400$ MeV on $64^3$ lattices is consistent with
those on $32^4$ lattices, however, more data are needed at $T> 300$
MeV.  Second, when fitting the data using the expected $c_0 + c_2/T^2
+ c_4/T^4$ form, the running of the QCD coupling requires the
asymptotic behavior to have a $g^4$ variation in $c_0$.  Current data
are not good enough to resolve this feature.

Overall, our preliminary results provide a reasonably
consistent picture of the EoS over the full temperature range and are
of sufficient precision to be incorporated into hydrodynamical models.
The current uncertainty of up to $20\%$ at $180-300$ MeV in the trace
anomaly is expected to reduce to roughly $5\%$ with inclusion of data
at $m_\ell/m_s = 0.05$ and new simulations on $N_\tau=12$ lattices.

\section{Entropy Density Across the Transition}

The energy and entropy densities are two very useful markers of the
crossover temperature as these are the quantities that enter into the
phenomenological analyses of heavy ion collisions.

\begin{figure}[ht]
  \begin{tabular}{cc}
    \includegraphics[width=.5\textwidth]{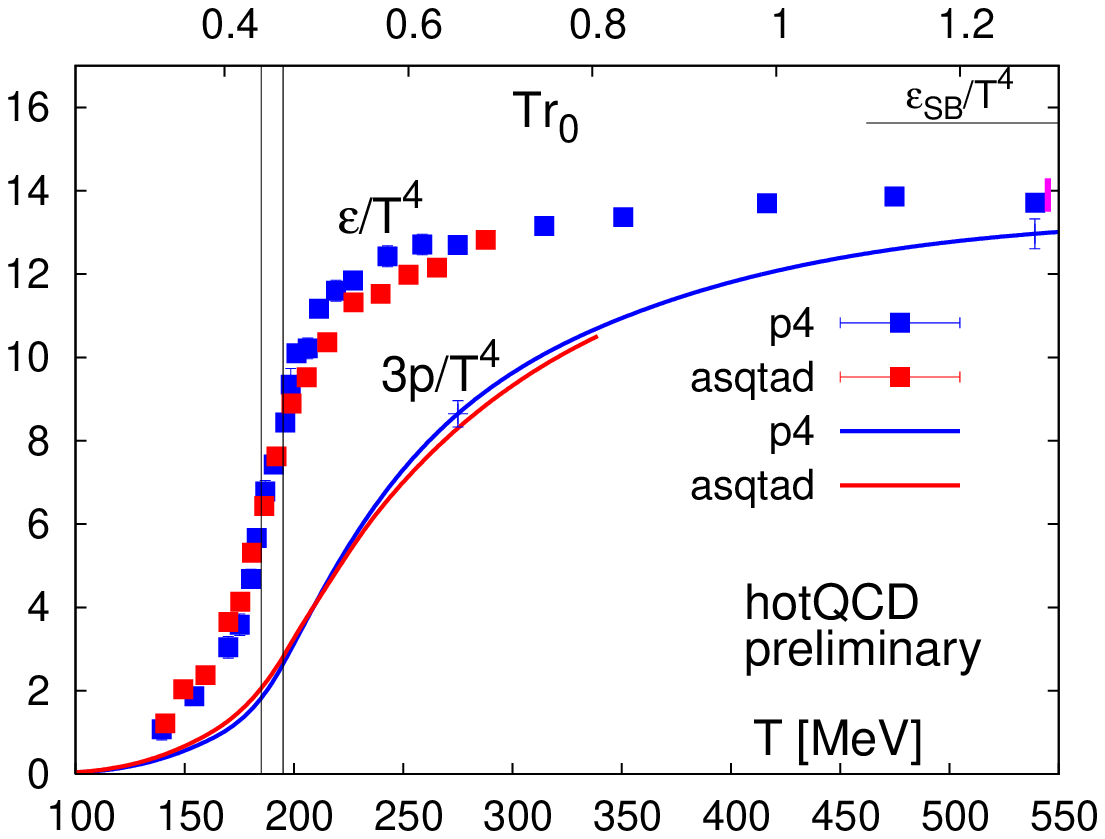}
    &
    \includegraphics[width=.5\textwidth]{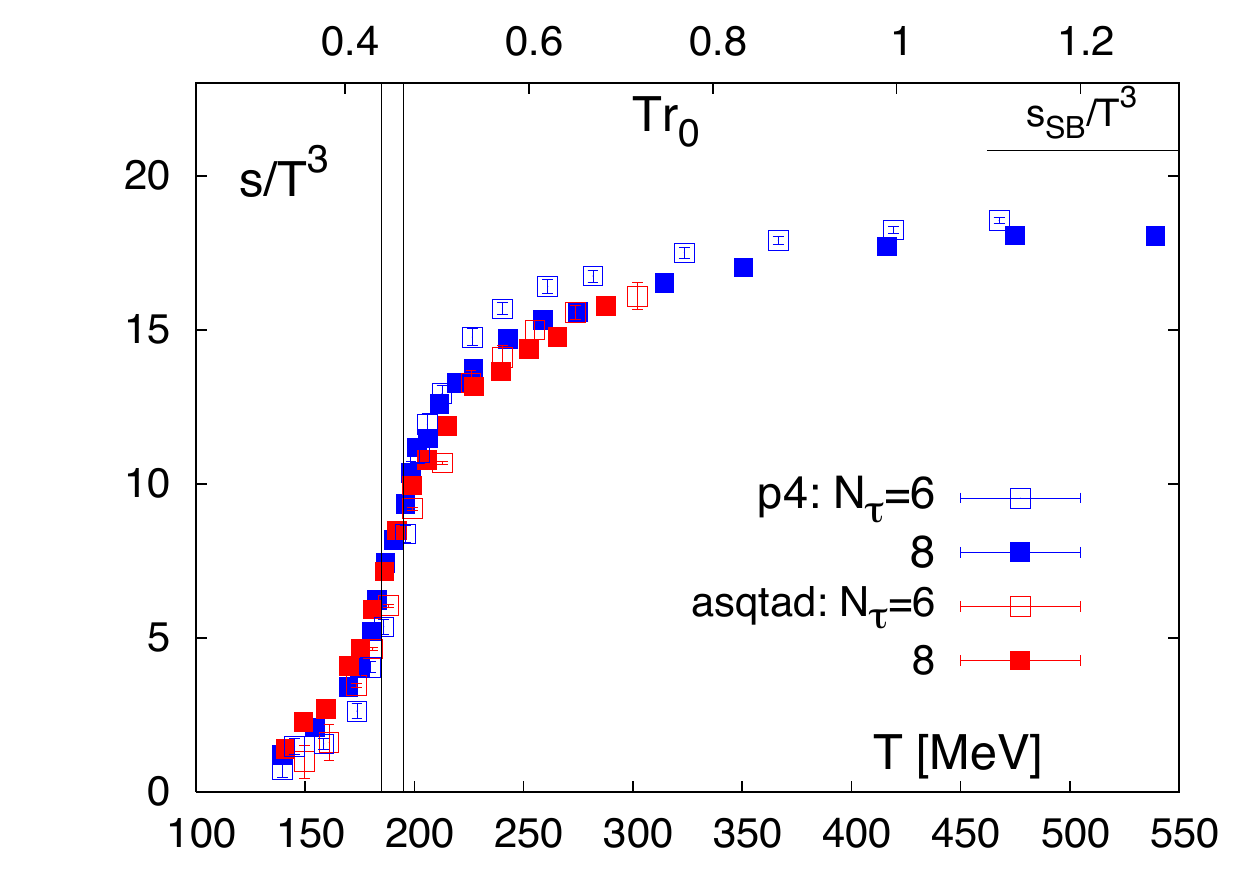}
  \end{tabular}
  \caption{Results for energy, pressure and entropy 
   density for the p4fat3 and asqtad
   actions with $m_\ell/m_s =0.1$ over $T=140-550$ MeV. The band $T=185-195$, 
   drawn to guide the eye, covers the inflection point. 
  \label{fig:entropy}
  }
\end{figure}

The data for energy density $\varepsilon$, pressure $p$, and the entropy
density $s/T^3 = (\varepsilon + p)/T^4$ are shown in
Fig.~\ref{fig:entropy}. We observe a rapid cross-over in the range
$175-205$ MeV and a smaller difference between $N_\tau=6$ and $8$, 
and conclude that quarks and gluons become the dominant and
relevant degrees of freedom above $250$ MeV, a region that will be
probed at the LHC.

\section{Deconfinement and Chiral Transitions}

This section summarizes results for quantities used to probe the
deconfinement (Polyakov loop and quark number susceptibility) and the
chiral (chiral condensate and its susceptibility) transitions. 

\begin{figure}[ht]
\begin{tabular}{cc}
    \includegraphics[width=.5\textwidth]{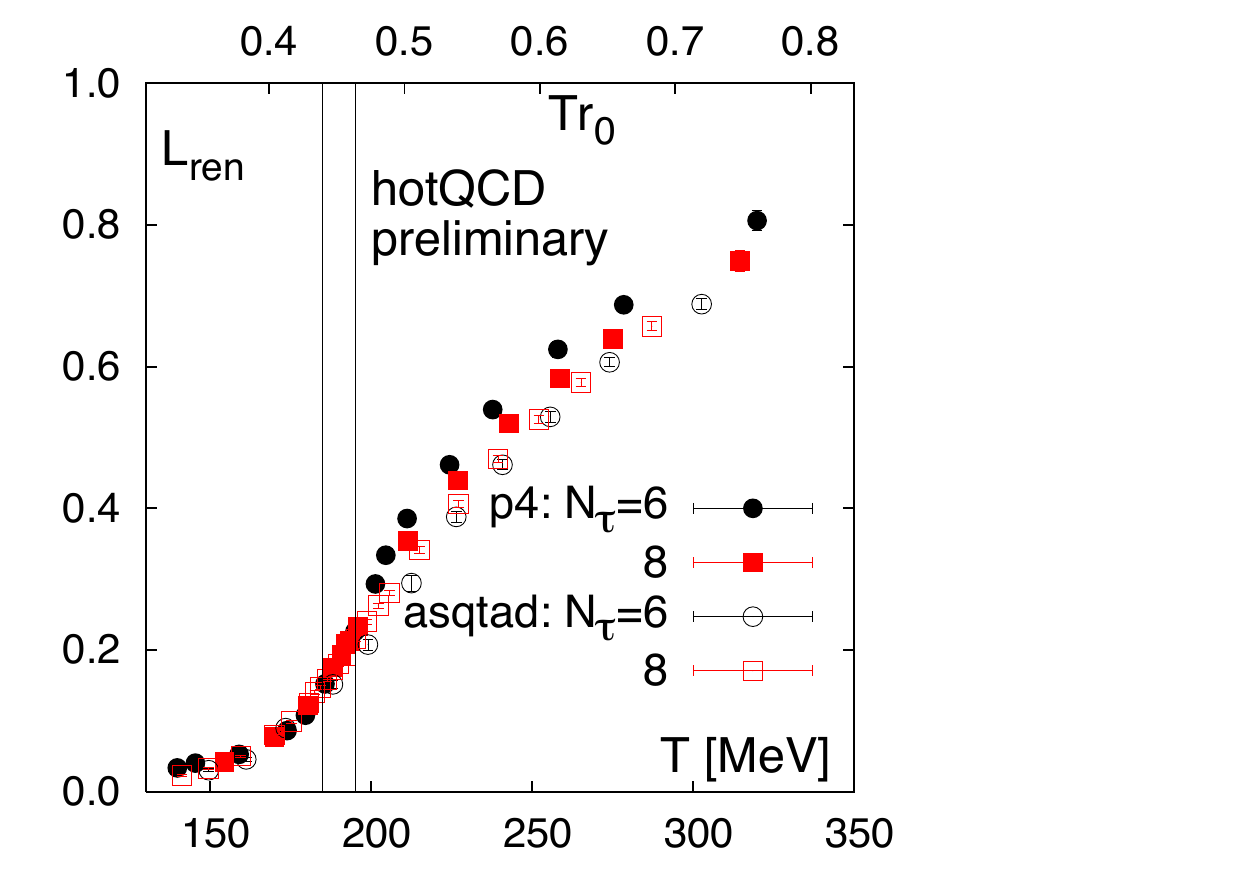}
&
    \includegraphics[width=.5\textwidth]{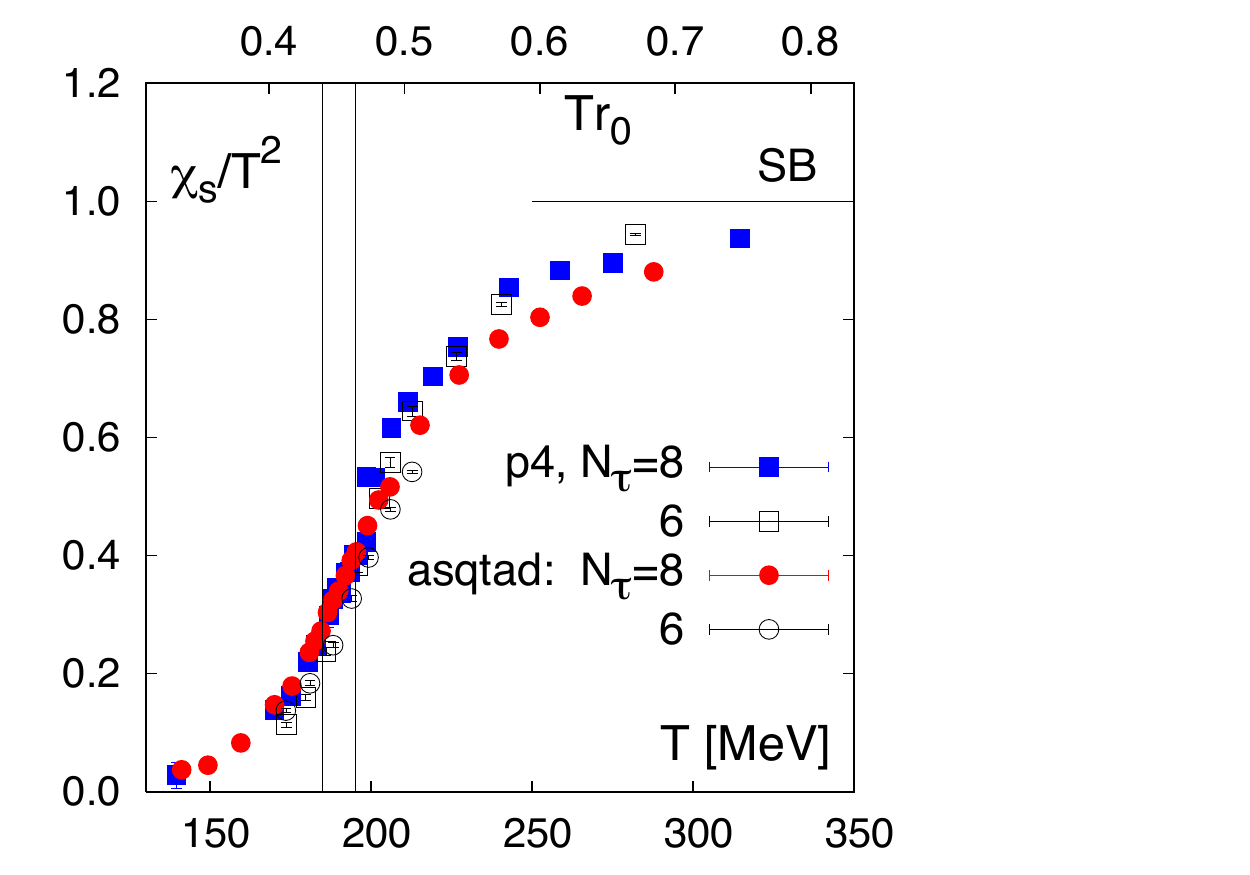}
\end{tabular}
  \caption{Deconfinement indicators for p4fat3 and asqtad
   actions. 
   (A) Renormalized Polyakov loop. 
   (B) Strange quark number susceptibility $\chi_s/T^2$. 
       The band $185-195$ MeV captures the inflection point. 
  \label{fig:Deconf}
  }
\end{figure}

The data for the renormalized Polyakov loop 
$\langle L_{\rm ren}(T)\rangle = Z(g^2)^{N_\tau} 
 \langle L_{\rm bare}(T) \rangle $,
which measures the free energy $F_\infty$ of an isolated quark,
$L_{\rm ren} = \exp[-F_{\infty}(T)/(T)]$, is shown in
Fig.~\ref{fig:Deconf}. There is a small, $< 10\%$, difference between
p4fat3 and asqtad data above $T=200$ MeV with the general trend that
the difference decreases with increasing $N_\tau$.  The continued slow
rise beyond $T \sim 250$ MeV results in a broad shoulder past the peak
in the associated susceptibility, and by $N_\tau=8$ there no longer is
a well defined inflection point (in $ L_{\rm ren}$) nor a peak in its
susceptibility. We find that locating the transition $T$ from the 
inflection point in $ L_{\rm ren}$ is already marginal at 
$N_\tau=6$~\cite{Cheng:2006}. 

P4fat3 and asqtad data for the strange quark number susceptibility,
$VT\chi_s = \partial {\rm ln}Z / \partial^2(\mu_s/T)$, are shown in
Fig.~\ref{fig:Deconf}. They are consistent for both $N_\tau=6$ and $8$
and show a rapid crossover with the inflection point covered by the
band at $185-195$ MeV.  $\chi_s$ is a good probe of deconfinement as
it measures fluctuations in the strange charge and therefore 
it does not need renormalization. The location of the peak in the
fluctuations, however, requires calculating the fourth derivative,
which is still in progress.

%% \subsection{Chiral symmetry restoration}

\begin{figure}[ht]
 \begin{center}
    \includegraphics[width=.3\textwidth]{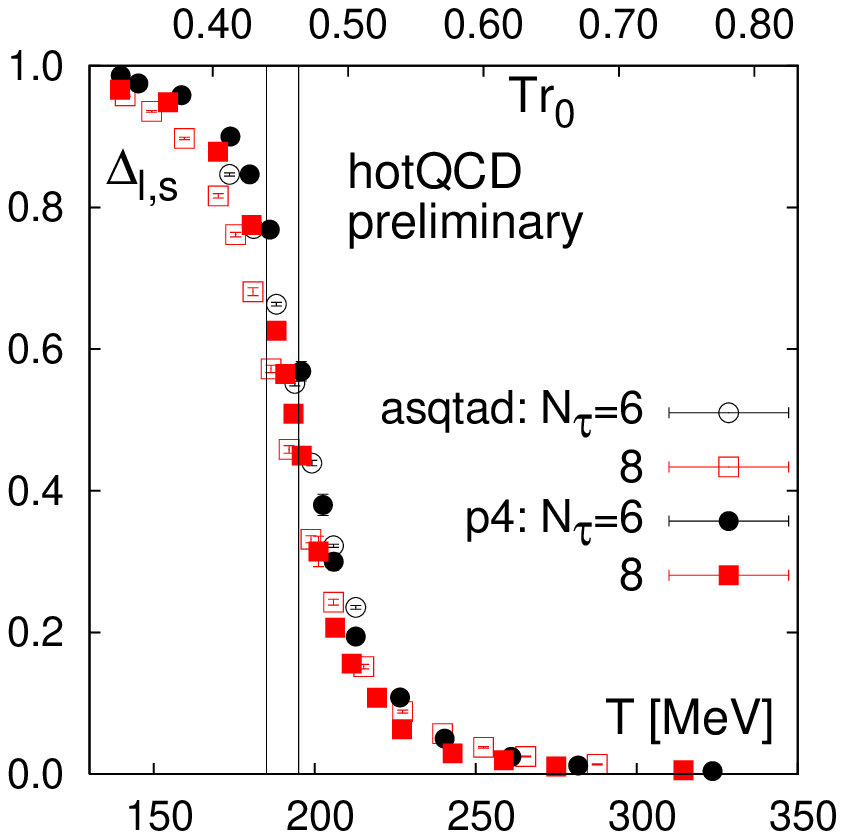}
    \hfill
    \includegraphics[width=.3\textwidth]{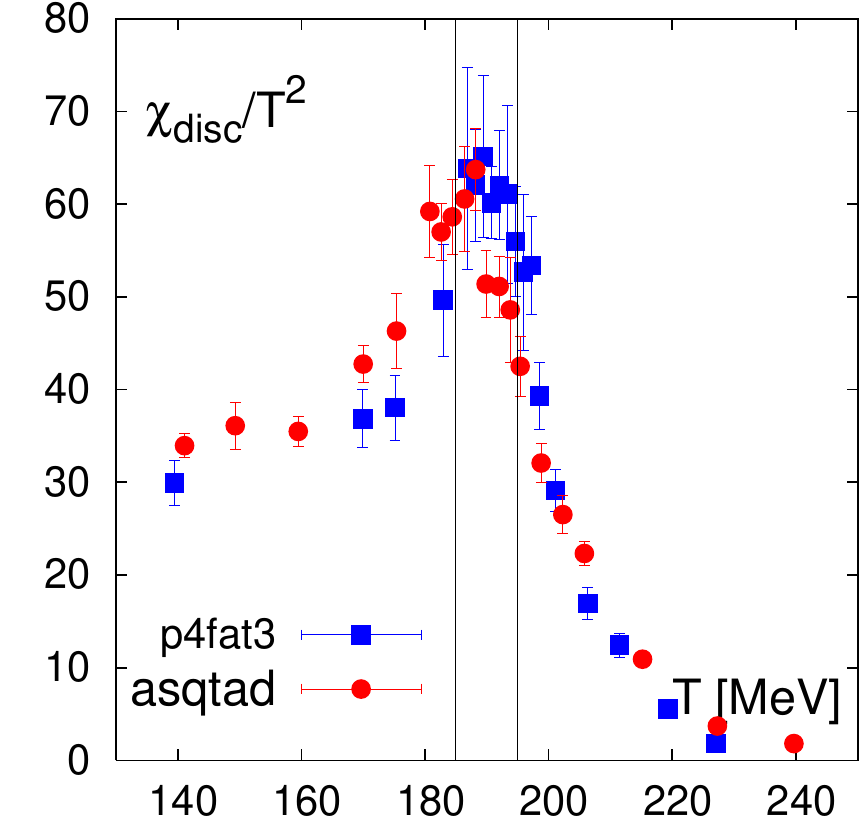}
    \hfill
    \includegraphics[width=.3\textwidth]{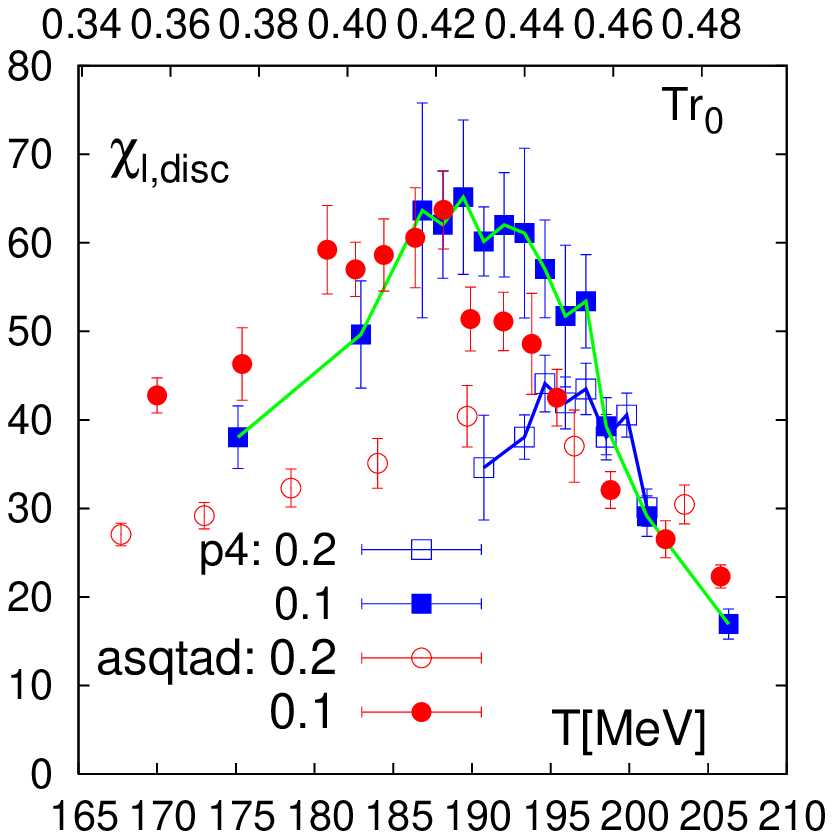} 
  \end{center}
  \caption{P4fat3 and asqtad data for the chiral symmetry restoration transition:
   (Left) The subtracted condensate $\Delta_{l,s}$; 
   (Center) $\chi_{disc}$ for $m_\ell = 0.1 m_s$;
   (Right) Comparison of peak in $\chi_{disc}$ with $m_\ell/m_s = 0.1$ and $0.2$.
    Note the broadening of the peak and strong $m_{quark}$ 
    dependence attributed to Goldstone modes ~\cite{Karsch:2008}.
  \label{fig:ChiDisconn}
  \label{fig:Delta}
  }
\end{figure}

The chiral condensate is investigated using the combination 
\begin{equation}
  \Delta_{l,s}(T) =  \frac{\pbp_\ell(T) - m_\ell/m_s \pbp_s(T)}
                          {\pbp_\ell(0) - m_\ell/m_s \pbp_s(0)} \\
\end{equation}
in which the additive ultraviolet divergence of the form $m/(a^2)$ at
nonzero quark mass is removed. The data in Fig.~\ref{fig:Delta} show
consistency between the two actions by $N_\tau=8$ indicating that the
residual different additive and multiplicative renormalization factors due to
slight differences in the lattice parameters between the two actions
mostly cancel in $ \Delta_{l,s} (T)$. The difference between
$N_\tau=6$ and $8$ data are mostly accounted for by shifting the $T$ of $N_\tau=6$ data
to the left by $\approx 5$ MeV.  Lastly, the inflection point in the crossover is again
captured by the band at $185-195$ MeV.

Fluctuations in the light quark condensate are given 
by the isosinglet chiral susceptibility. It consists of the 
connected and disconnected parts  
$ \chi_{singlet} = \chi_{\rm disc} + 2 \chi_{\rm conn} $.
Data for $\chi_{\rm disc}$ for the two actions on $N_\tau=8$ lattices
are compared in Fig.~\ref{fig:ChiDisconn} and exhibit four
features. (A) Consistency between the two actions; (B) an increase in
peak height with decreasing quark mass; (C) A shift by $\sim 10$ MeV
towards smaller $T$ between $m_\ell/m_s = 0.2$ and $0.1$; and (D) a broadening of the peak
as $m_q \rightarrow 0$. Karsch~\cite{Karsch:2008} has argued that,
below the transition in the broken chiral symmetry phase,
fluctuations caused by the vanishing Goldstone pion mass 
should broaden the peak, $i.e.$, 
$\chi_{singlet} \to \infty$ with $m_q \to 0$ for all $T \le T_c$. If this picture 
(consistent with our data) is substantiated
then the location of the right edge of this
broad peak would be the appropriate locator of the chiral symmetry
restoration temperature.  Extrapolating this edge, 
using our $m_\ell/m_s =0.1$ and $0.2$ data, to the chiral limit 
gives $T_{transition} > 180$ MeV for $N_\tau=8$.

\section{Conclusions}

Data for $(\varepsilon-3p)/T^4$ show consistency between asqtad and
p4fat3 actions and exhibit $< 20\%$ change between $N_\tau = 6$ and
$8$. We consider estimates of $\varepsilon$ and $p$ on $N_\tau=8$ lattices 
are precise enough to use in phenomenological analyses
of the evolution of the QGP at RHIC ($T < 300$ MeV).

All observables used to probe the transition (energy and entropy
density, Polyakov loop, quark number susceptibility, chiral condensate
and its susceptibility) show a rapid crossover that takes place
between $175 \leq T \leq 205$ MeV with $185-195$ MeV as our best
estimate for the transition temperature most relevant for
phenomenological studies.

These calculations are being extended in two ways. First, simulations
are being done at $m_\ell = 0.05 m_s$ to directly probe the system at
approximately the physical $u,d$ mass~\cite{Soeldner:2008}. Second, for 
continuum extrapolation we plan to simulate $N_\tau = 12$ lattices in addition to 
$N_\tau=6$ and $8$.

\vspace{-1mm}
\acknowledgments 

We are grateful to LLNL, NNSA, and New York Center for
Computational Science for providing access to the Bluegene/L
supercomputers.  This work is supported by US DOE and NSF.

\end{document}